\documentclass[aps,prl,showpacs,twocolumn]{revtex4-2}
\usepackage{graphicx}
\usepackage{subfig}
\usepackage{tabularx}
\usepackage{bm}
\usepackage{amsmath}
\usepackage{esvect}

\usepackage{hyperref}
\usepackage[capitalize]{cleveref}

\providecommand{\ket}[1]{\lvert #1 \rangle}

\providecommand{\be}{\begin{equation}}
\providecommand{\ee}{\end{equation}}
\providecommand{\ba}{\begin{eqnarray}}
\providecommand{\ea}{\end{eqnarray}}

\usepackage{mathbbol}

\usepackage{tikz}
\usepackage{amsfonts,amssymb}
\usepackage{dsfont}
\usepackage{physics}
\usepackage{tocvsec2}

\newcommand{\beq}{\begin{equation}}
\newcommand{\eeq}{\end{equation}}

\hypersetup{colorlinks=true,linkcolor=blue,citecolor = blue, urlcolor=black}

\usepackage{appendix}

\begin{document}

\title{Gottesman-Kitaev-Preskill encoding in continuous modal variables of single photons}

\author{ Éloi Descamps$^{1}$}
\author{ Arne Keller$^{1,2}$}
\author{Pérola Milman$^{1}$ }
\email{corresponding author: perola.milman@u-paris.fr}

\affiliation{$^{1}$Laboratoire Mat\'eriaux et Ph\'enom\`enes Quantiques, Universit\'e Paris Cité, CNRS UMR 7162, 75013, Paris, France}
\affiliation{$^{2}$Departement de Physique, Université Paris-Saclay, 91405 Orsay Cedex, France}

\begin{abstract}

GKP states, introduced by Gottesman, Kitaev, and Preskill, are continuous variable logical qubits that can be corrected for errors caused by phase space displacements. Their experimental realization is challenging, in particular using propagating fields, where quantum information is encoded in the quadratures of the electromagnetic field. However, travelling photons are essential in many applications of GKP codes involving the long-distance transmission of quantum information. We introduce a new method for encoding GKP states in propagating fields using single photons, each occupying a distinct auxiliary mode given by the propagation direction.  The GKP states are defined as highly correlated states described by collective continuous modes, as time and frequency. We analyze how the error detection and correction protocol scales with the total photon number and the spectral width. We show that the obtained code can be corrected for displacements in time-frequency phase space - which correspond to dephasing, or rotations, in the quadrature phase space - and to photon losses. Most importantly, we show that generating two-photon GKP states is relatively simple, and that such states are currently produced and manipulated in several photonic platforms where frequency and time-bin biphoton entangled states can be engineered. 

\end{abstract}
\vskip2pc 
 
\maketitle

Derived from classical error correction protocols, quantum error correction plays a central role in quantum information theory. The counterintuitive features of quantum mechanics are inherently fragile and necessitate error correction to enable the manifestation of the quantum advantage of protocols over their classical counterparts. Originally devised for qubits and finite-dimensional discrete systems \cite{Shor,Steane}, quantum error-correcting codes rely on creating redundant states composed of multiple physical qubits to define logical qubits. By measuring well chosen observables that do not affect the state of the logical qubit - the stabilizers - one can detect and correct errors affecting the physical qubits. Entanglement is usually an important ingredient in error correction, as exemplified by the states that withstand physical qubit flips and dephasing \cite{5qubits,gottesman1997stabilizer}.

When dealing with continuous variables such as position and momentum, it is also possible to encode quantum information into states that can be corrected for errors \cite{PhysRevA.56.1114,PhysRevA.59.2631,PhysRevLett.80.4084,PhysRevLett.80.4088}. One of the most successful model for error correction in continuous variables is known as the GKP code \cite{gottesman_encoding_2001} (see, for instance, \cite{PRXQuantum.2.020101} for a review), named after its creators: Gottesman, Kitaev, and Preskill. GKP states are a direct extension of the discrete variable codes into the continuous domain \cite{PhysRevLett.78.405,PhysRevA.54.1862} and are correctable for errors modeled as displacements in phase space. In the realm of quantum optics, one usually thinks of encoding GKP states using two orthogonal quadratures of the electromagnetic field, since the associated observables obey the same commutation relation as position and momentum. In such a system, phase space displacements can be the consequence of unwanted interference with a parasite classical field or model photon losses \cite{albert_performance_2018, Joshi_2021}. However, a main difficulty consists of the experimental production of such highly non-classical states, which involves, for instance, the prior (non-deterministic) production of Schrödinger cat-like states (or, in practice, kittens) that are made to interfere \cite{PhysRevA.101.032315}. Several proposals exist \cite{PhysRevA.102.062411, Bourassa2021blueprintscalable,PRXQuantum.2.030325,PhysRevA.108.012603, Eaton_2019, PhysRevX.13.031001}, as well as a first experimental realization \cite{konno2023propagating}. Of course, one can also encode GKP states using other bosonic systems, as superconducting circuits \cite{GKPSupra} or the motional states of trapped ions \cite{Ions, TrappedionsGKP}. However, building a robust code adapted to propagating fields is clearly of major importance if one wills to transmit quantum information, in particular, to several independent users \cite{SaraEleni}. 

In the present Letter we propose a new way to encode GKP states in quantum optics using an original approach to continuous variables. We use continuous collective variables of single photons occupying distinct auxiliary modes, as the propagation direction \cite{PhysRevLett.131.030801}, to define redundant states. Our model can apply to different single photon continuous modes, as time and frequency (that we discuss in detail), the transverse position and momentum \cite{tasca_continuous_2011}, the propagation direction   \cite{PhysRevX.4.031007,Raymond:23} and also to the collective modes of massive particles, as the normal modes of trapped ions \cite{Normal1, Normal2}. Using the established formalism, we theoretically investigate how GKP codewords can be defined in a system comprising $n$ individual photons, elucidating how some known properties of the encoded states can be retrieved, as for instance the scaling of the error rate with the number of photons, their possibility to recover from photon losses and the effects of imperfect preparation and measurement. Notably, we show that in the time-frequency (TF) encoding scheme, these properties have a fundamentally different physical origin compared to quadrature-based (QB) encoding. Finally, we demonstrate that the generation and manipulation of TF GKP states in quantum optics is already a reality in laboratories. This is particularly evident in experimental setups where entangled photon pairs with a correlated comb-like temporal or spatial structure are observed, as exemplified in references \cite{PhysRevA.102.012607, Qudits, Comb, Imany:18, PhysRevA.82.013804, massive, PhysRevA.95.042311, 10.1063/5.0089313, SaraEleni,Lukens:17, AndreaSabattoli:22, Zhang:18, NatureBins, Kaneda:19}. Consequently, these experimental platforms and the associated quantum protocols can immediately benefit from our findings. 

We define as ${\cal S}_n$ the subspace consisting of $n$ single photons occupying each an auxiliary mode, as the propagation direction. Photons are characterized by a collective spectral function that also depends on the auxiliary mode. The auxiliary modes can be seen as external degrees of freedom and frequency as internal degrees of freedom \cite{PhysRevX.11.031041}. Pure states are written as 
\be\label{state}
\ket{\psi}=\int d\omega_1...d\omega_n f(\omega_1,...,\omega_n)\ket{\omega_1,...,\omega_n},
\ee 
where $f$ is a normalized function (the spectral amplitude) that determines the properties of state \eqref{state}, as entanglement and its mode decomposition \cite{fabre_modes_2020, PhysRevLett.84.5304}, and $\hat a_i^{\dagger}(\omega_i)\ket{0}=\ket{\omega_i}$. In ${\cal S}_n$, errors are represented in the basis of time and frequency displacements, and they can affect the photons locally, {\it i.e.}, they act independently on the photons of each auxiliary mode, a situation similar to the one affecting a collection of physical qubits. Such displacements are described by operators acting on a given mode $j \in \{1,...,n\}$ as $\hat D_{\hat \omega_j}(\delta_{t_j})=e^{-i \hat \omega_j \delta_{t_j} }$ and $\hat D_{\hat t_j}(\delta_{{\omega}_j})=e^{-i \hat t_j \delta_{{\omega}_j} }$, where $\hat \omega_i=\int d\omega \omega \hat a_i^{\dagger}(\omega)\hat a_i(\omega)$ and $\hat t_i = \int dt t \tilde {\hat a}_i^{\dagger}(t)\tilde {\hat a}_i(t)$,  with $\tilde {\hat a}_i(t)=\frac{1}{\sqrt{2\pi}}\int d\omega e^{i\omega t}\hat a(\omega)$, where $\hat a_i^{\dagger}(\eta)$ creates one photon at frequency $\eta$ at the $i$-th auxiliary mode and $[\hat \omega_k,\hat t_j ]=i\delta_{k,j}\int d\omega \hat a_k^{\dagger}(\omega)\hat a_k(\omega)=\hat n_k i\delta_{k,j}$ ($\hat n_k = \mathbb{1}$ on ${\cal S}_n$) \cite{PhysRevA.105.052429, PhysRevLett.131.030801}.  
Now we show that entangled states of $n$ photons in the TF continuous variables sharing the same properties of GKP states can be corrected for this type of errors. Such states can be written in the general separable form in collective variables:
\be\label{entangled}
\ket{\overline k}=\int d\Omega_1...d\Omega_n F_k (\Omega_1)\Pi_{i>1}^nG_i(\Omega_i)\ket{\omega_1,...,\omega_n},
\ee
where $\Omega_j=\sum_i^n \alpha_{i,j} \omega_i$ are collective variables, $\alpha_{i,j}$ is an invertible matrix with $\alpha_{i,j} \in \{-1/\sqrt{n},1/\sqrt{n}\}$ and $k \in \{0,1\}$. We consider for simplicity and without loss of generality that $\alpha_{i,1}=1/\sqrt{n}~\forall~i$, and that $n = 2^m$, $m \in  \mathbb {N}$. The matrix $\alpha$ is unitary and symmetric, hence
$\omega_i = \sum_j \alpha_{j,i}\Omega_j$.

An ideal $n$ photon GKP state $\ket{\overline k}$ in ${\cal S}_n$ can be defined from \eqref{entangled} using $F_{k}(\Omega_1)=\sum_{s=-\infty}^{s=\infty}\delta(\Omega_1-(2s+k)\Omega_o)$, where $\delta$ is the Dirac delta function and $\Omega_o$ is an arbitrary (constant) frequency. Hence, the logical qubits $\ket{\overline 0(\overline 1)}$ are non-physical states formed by an infinity of peaks localized at frequencies which are integer multiples of $\Omega_o$, and $2\Omega_o$ is the peak interspacing in each logical qubit (the choice of $\alpha_{i,j}$ means that we have supposed that all the photons' frequencies are equally spaced \cite{Note2}). States $\ket{\overline k}$ are defined uniquely using the collective variable $\Omega_1$. The functions $G_i$ are arbitrary and their role, not crucial for the code working principles, will be discussed later in this manuscript. Thus, all the relevant information for error diagnosis and correction is contained only in variable $\Omega_1$, and we disregard the information contained in $\Omega_{i>1}$. This type of situation is current in quantum optics where different physical properties, as group and phase velocity for multi-modal fields, are associated to different collective variables. For instance in the Hong-Ou-Mandel (HOM) experiment \cite{PhysRevLett.59.2044} (see \cite{doi:10.1142/S0217979207038186, douce_direct_2013} for its generalization to many photons), the variable $\Omega_1=(\omega_1-\omega_2)/\sqrt{2}$ is directly measured, while the information in variable $\Omega_2=(\omega_1+\omega_2)/\sqrt{2}$ is  disregarded \cite{NoteHOM}. By combining different interferometric techniques \cite{sergienko}, one can access different collective variables measuring not only frequency but other continuous modes, as the transverse position and momentum \cite{tasca_continuous_2011, PhysRevA.95.042311}. 

TF GKP states are intrinsically multimode states relying on the particle-mode non-separability, so they are fundamentally different from optical combs in single mode states using spectral engineering of classical (coherent) states or single photons \cite{PhysRevA.102.012607, PhysRevLett.130.200602}. Thanks to the encoding in collective variables (modes) of individual photons, TF GKP states reveal their multi-photonic properties, as the scaling of the error tolerance with the number of photons $n$. Hence, they're also fundamentally distinct from QB GKP states - that can be defined in single modes -, even in their multi-dimensional version \cite{PRXQuantum.3.010335}. 

An example of a possible $\ket{\overline k}$ state with $G_i(\Omega_i)= \delta(\Omega_i) ~\forall ~i$ in \eqref{entangled} is \cite{MacconeNature}:
\be\label{GKP}
\ket{\overline k}=\sum_{s=-\infty}^{\infty}\ket{(2s+k)\frac{\Omega_o}{\sqrt{n}}}_1...\ket{(2s+k)\frac{\Omega_o}{\sqrt{n}}}_n,
\ee 
Analogously to the QB GKP states, we can identify non-Hermitian operators that act in $\ket{\overline k}$ as Pauli matrices \cite{PhysRevA.102.012607}. One way to see this is using displacements in the {\it collective} TF variables, $\hat D_{\hat T_1}(\Delta_{\omega})=e^{-i  \hat T_1 \Delta_{\omega}}$ and $\hat D_{\hat \Omega_1}(\Delta_t)=e^{-i \hat \Omega_1 \Delta_t }$, where ${\Delta_{t(\omega)}} \in \mathbb{R}$, $\hat \Omega_1= \sum_i^n\hat \omega_i/\sqrt{n}$ and $\hat T_1 = \sum_i^n \hat t_i/\sqrt{n}$, with $[\hat \Omega_1,\hat T_1 ]=\mathbb{1}i$ in ${\cal S}_n$ \cite{PhysRevA.105.052429}. Collective operators can also be associated to variables $\Omega_{i> 1}$. By an appropriate choice of ${\Delta_{t(\omega)}}$, we can define the Pauli-like operators in the TF GKP subspace as $\hat X=e^{-i \hat \Omega_1  T_o}$ and $\hat Z=e^{-i \hat T_1 \Omega_o }$, with $T_o=\pi/\Omega_o$ and  $\hat Y=i\hat Z \hat X$ so $\ket{\overline 1}=\hat X\ket{\overline 0}$. Also, a combination of the universal time-frequency gates defined and physically described in \cite{PhysRevA.105.052429}, can be used to complete the universal gate-set in the TF GKP space \cite{SM}. A key aspect of the proposed encoding is that the construction of the logical operators $\hat X$, $\hat Y$ and $\hat Z$ (and consequently of the TF GKP universal gate-set) is not unique.  
\begin{figure}[h]
    \includegraphics[width=\columnwidth]{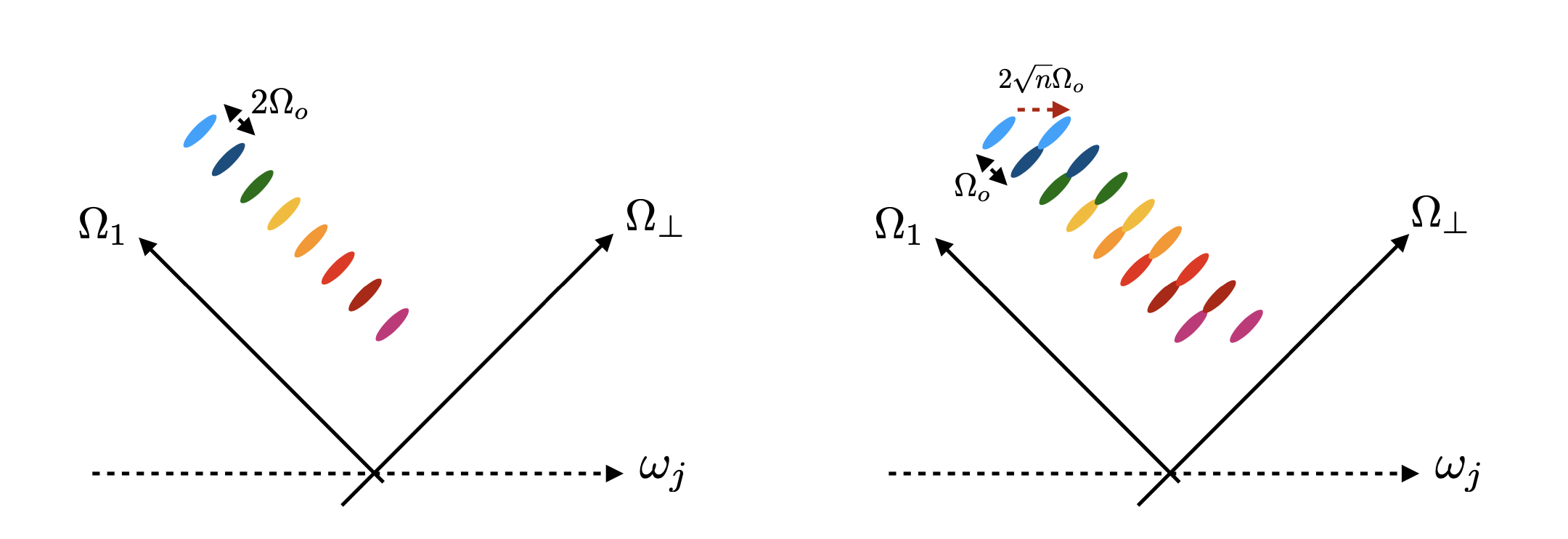}
    \caption{Left: Physical TF GKP state with peak spacing $2\Omega_o$ in the collective variable $\Omega_1$. $\Omega_{\perp}$ is a collective variable orthogonal to $\Omega_1$. Right: A displacement of $2\sqrt{n}\Omega_o$ in the local variable $\omega_j$ also displaces variable $\Omega_1$. The displacement in the orthogonal direction $\Omega_{\perp}$ is not relevant since we ignore the information it contains. }
  \label{fig1}
\end{figure}
Nevertheless, the formal construction of the TF GKP code is identical to the QB one, so all the properties of the latter can be retrieved here, but now associated to different continuous variables: states \eqref{GKP} enable correcting for {\it collective} time and frequency errors corresponding to TF displacements such that $|\Delta_{\omega} |<\Omega_o/2$ and $|\Delta_t|<\pi/(2\Omega_o)$, delimiting a TF phase space area of correctable errors satisfying $4|\Delta_{\omega}||\Delta_t|<\pi$. Physically, displacement errors can correspond to imperfect aligning of an interferometer and the effects of a non-linear device as an optical fiber. 

Restricting to collective errors is missing our main goal, which is creating states which are robust against {\it local} displacements in the TF variables of each photon. Physically, this corresponds to a situation where each photon occupying a different propagation mode is distributed throught different channels and frequency and time displacements occur independently in each mode (and consequently each photon). Such TF displacements are rotations in the quadrature phase space, a type of error against which the usual QB GKP are not very efficient \cite{PRXQuantum.2.020101} and require using rotation symmetric states in the quadrature space \cite{PhysRevX.10.011058}. As for states \eqref{GKP}, they are robust against global and local rotations by construction. We now study the effect of local noise in the TF GKP states \eqref{entangled}. We have:
\begin{eqnarray}\label{displace}
&&\hat D_{\hat \omega_j}(\delta_{t_j})\ket{\overline k}= \\
&&\int d\Omega_1...d\Omega_n e^{-i\frac{\Omega_1\delta_{t_j}}{\sqrt{n}}}F_k (\Omega_1)\prod \limits_{k>1}^n \tilde G_k(\Omega_k)\ket{\omega_1,...,\omega_n},\nonumber
\end{eqnarray}
where $ \tilde G_k(\Omega_k)= e^{-i \frac{\alpha_{j,k}\Omega_k \delta_{t_j}}{\sqrt{n}}} G_k(\Omega_k)$.  In addition,
\begin{eqnarray}\label{displace2}
&&\hat D_{\hat t_j}(\delta_{\omega_j})\ket{\overline k}=\\
&&\int d\Omega_1...d\Omega_n F_k \left (\Omega_1-\frac{\delta_{\omega_j}}{\sqrt{n}}\right )\prod \limits_{k>1}^n G_k(\Omega_k-\bar \Omega_k)\ket{\omega_1,...,\omega_n}, \nonumber
\end{eqnarray}
where $\bar \Omega_k = \alpha_{j,k}\delta_{\omega_j}$. Eqs. \eqref{displace} and  \eqref{displace2} lead to an important result: the codewords \eqref{GKP} protect against shifts in local variables $\omega_j$ in a way that scales with $\sqrt{n}$ with the number of photons $n$. Some examples of correctable errors are then: a single photon in mode $j$ that is displaced by $|\delta_{t_j}| < \sqrt{n}\pi/(2\Omega_o)$; up to $\approx \sqrt{n}$ photons in different modes $j$ that are each displaced by $|\delta_{t_j}| < \pi/(2\Omega_o)$;  $n$ photons that are each equally displaced by $|\delta_{t_j}| < \pi/(2\Omega_o\sqrt{n})$. Thus, if one focus on {\it local} errors, the codewords \eqref{GKP} can protect for them provided that they lie in a phase space area of size $ 4|\sum_j^n \delta_{\omega_j}||\sum_j^n\delta_{t_j}|<n\pi$ \cite{PhysRevA.97.032346}. We can see this as a re-scaling of the code, since the overall protection corresponds to the one of a single photon GKP states formed by peaks that are distant of $2\sqrt{n}\Omega_o$ and $\sqrt{n}/(2\Omega_o)$ in time and frequency variables, respectively. Interestingly, contrary to what one would observe in a classical time frequency Fourier relation where the dilatation of the frequency space is accompanied by the shrinking of the time space and vice-versa, the observed effective phase space dilatation is a geometric consequence of encoding information in collective variables while errors occur in local ones. This re-scaling leads to a photon number dependency of the probability error rate analogous to the one observed for QB GKP encoding, as we'll see later. 

Operators $\hat X$ and $\hat Z$ are not unique. We can define $\hat X_{j}=e^{-i\hat \omega_j T_o\sqrt{n}}$ and check by computing $\hat X_j\ket{\bar k}$ that $\hat X_{j}$ acts in variables $\Omega_1$ in the same way as $\hat X$ does (see Fig. \ref{fig1} and \cite{SM}). Using this, we can detect the loss of one photon in mode $j'$ (unknown) and adapt to its effects by measuring time and frequency displacements only. If a photon is lost, the TF GKP state becomes $\ket{\bar k}_{-1}=\int d\omega \hat a_{j}(\omega)\ket{\bar k}=\hat {\cal E}_j\ket{\bar k}$ (we considered that the photon loss rate is independent of the frequency \cite{PhysRevA.31.3761}). Defining $\hat S_j = e^{-i \eta_j\Omega_o\hat t_j}\hat X_{j}^2 e^{i \eta_j\Omega_o\hat t_j}=e^{-i2(\hat \omega_j-\hat n_j\eta_j\Omega_o) T_o\sqrt{n}} $, $\eta_j \in \mathbb{R}$, and only considering the information in $\Omega_1$, we have that $\hat S_{j} \hat S_{j+1}$ stabilizes $\ket{\overline k}$ in ${\cal S}_n$ if $(\eta_j+\eta_{j+1})\sqrt{n}=m$, $m \in \mathbb{Z}$, for all $j$, and it stabilizes $\hat {\cal E}_{j'}\ket{\bar k}=\ket{\bar k}_{-1}$ if $j' \neq j, j+1$. In addition, $\hat S_j \hat S_{j+1}\hat {\cal E}_j\ket{\overline k}=\hat {\cal E}_j\hat S_{j+1}\ket{\overline k}=e^{-2i\eta_j \pi\sqrt{n}}\hat {\cal E}_j\ket{\overline k} $, so by judiciously choosing $\eta_j$ we can detect a photon loss and the mode from which it was lost ($j$ or $j+1$ here). We can also define $\hat S= \Pi_j \hat S_j$, which is a stabilizer of $\ket{\overline k}$. Using that $\hat S \hat {\cal E}_j \ket{\overline k}=e^{-i\eta_j\pi\sqrt{n}}\hat {\cal E}_j \ket{\overline k}$ we can detect in a single shot that a photon has been lost and from which mode, by judiciously choosing $\eta_j$'s and if $\sum_{j=1}^n \eta_j \sqrt{n}= m$ \cite{SM}. The stabilizer measurement provides essential information about the mode that lost a photon, permitting to adapt the operations and measurements to a $n-1$ photon configuration: if a photon is lost, the collective effects in displacements are smaller, and in order to have $e^{2i\hat \Omega_1 T_o'} \ket{\overline k}_{-1}=\ket{\overline k}_{-1}$ we must use $T_o'= T_o n/(n-1)$. One can also re-insert the lost photon by applying a two-photon conditional gate involving mode $j$ and an arbitrary mode $i'$ in the code, and a displacement, so that $\hat D_{\hat T_1}(-\omega_j)e^{i\hat t_j \hat \omega_{i'}}\hat a_j^{\dagger}(\omega_j)\ket{\overline k}_{-1}=\ket{\overline k}$ \cite{SM, Jeannic, PhysRevA.105.052429}.

States \eqref{GKP} are not physical. We can define their normalizable version, $\ket{\tilde 0(\tilde 1)}$, using $\tilde F_k(\Omega_1)=\sum_{s=-\infty}^{s=\infty} e^{-\kappa^2 (2s+k)^2\Omega_o^2}e^{-\frac{(\Omega_1-(2s+k)\Omega_o)^2}{\Delta^2}}$, $k \in \{0,1\}$, where each peak of the TF GKP code has a Gaussian spectrum of width $\Delta (\ll \Omega_o$) in variable $\Omega_1$ and the comb of peaks distribution is modulated by a Gaussian envelope of width $\kappa^{-1} \ll T_o/\pi$ \cite{gottesman_encoding_2001}. We'll consider for simplicity that $\Delta = \kappa$. A finite width provides an intrinsic error probability to states $\ket{\tilde 0(\tilde 1)}$,  seen as perfect states \eqref{GKP} that have been subjected to a distribution of displacements (errors). Hence, an error probability ${\cal E}(\Delta/\Omega_o)$ is  associated to the error correction protocol through the definition of non-perfectly orthogonal states as codewords. The finite spectral width can be modeled as independent displacements of individual photons with a Gaussian amplitude distribution of width $\Delta_j$. Each photon $j$ in \eqref{GKP} is described by state 
\begin{eqnarray}\label{errorphoton}
&&\ket{\widetilde {(2s+k)\frac{\Omega_o}{\sqrt{n}}}}_j= \\
&&\int d\omega e^{-\Delta_j^2 (2s+k)^2\left (\frac{\Omega_o}{\sqrt{n}}\right)^2}e^{-\frac{\left (\omega-\frac{(2s+k)\Omega_o}{\sqrt{n}}\right )^2}{\Delta_j^2}}\ket{\omega}_j.\nonumber
\end{eqnarray}
The local variables $\omega_i$ behave as independent random variables, and we can suppose $\Delta_j=\Delta$. Each $n$ photon peak has the form $\Pi_{j}^n \ket{\widetilde {(2s+k)\frac{\Omega_o}{\sqrt{n}}}}_j$, and the total state's temporal envelope is $\Delta^{-1}$. By changing to the collective variables $\Omega_i$, each peak is described by a Gaussian spectral distribution of width $\Delta$ in all variables $\Omega_{i}$, leading to $\tilde F_k(\Omega_1)$ shown above. Hence, differently from the QB GKP encoding, the spectral width does not depend on the average photon number, and the probability of mistaking $\ket{\tilde 0}$ and $\ket{\tilde 1}$ is given by ${\cal E}(\Delta/\Omega_o)= \frac{\Delta}{\pi\Omega_o}e^{\frac{-\pi \Omega_o^2}{4\Delta^2}}$ \cite{gottesman_encoding_2001}. We recall however that in the present encoding the {\it effective} displacement on the collective variable $\Omega_1$ decreases with $\sqrt{n}$, increasing the effective distance between peaks. Consequently, the region of potential overlap between the two supposedly orthogonal QB GKP qubit states is modified \cite{CommentDisp}. 

By analyzing the effects of photon losses discussed above for physical TF GKP states, we see that the loss of $m$ photons will not significantly affect the code if the cumulated effective peak interspacing modification $T_o m/n$ \cite{SM} in the whole state ($\approx 1/(2\Delta T_o)$ peaks) is within each peak's half-width $\Delta/2$, leading to the condition $2T_o m/n\times 1/(2\Delta T_o) = m/(\Delta n) \leq \Delta$, or $n/m\geq 1/\Delta^2 \gg \pi/(\Omega_o T_o)=1$. These errors propagate with the number of gates, and a detailed analysis should be carried, but displacement based correction strategies can be devised based on this scaling \cite{SM}.  

We now discuss the role of a finite frequency width in the collective variables $\Omega_{j> 1}$. For simplicity, we'll consider $\Omega_\perp$ to be one of these variables and ignore all the others, considering a state with spectral amplitude $F_k(\Omega_1)G(\Omega_\perp)$. The width of the spectral distributions $F_k$ and $G$ in \eqref{entangled} can be independent and related to different physical constraints, as for instance energy conservation and the phase matching condition in spontaneous parametric down-conversion (SPDC). If the spectral function of state \eqref{entangled} is separable in variables $\Omega_1$ and $\Omega_{\perp}$ and all the measurements performed on variable $\Omega_1$, the spectral width $\sigma$ or the particular shape of $G$ have no importance. However, state preparation may be imperfect, leading to a state that is still separable but in the variables $\Omega_1'=\cos \theta \Omega_1 + \sin \theta \Omega_{\perp}$ and $\Omega_{\perp}'=\cos \theta \Omega_{\perp} - \sin \theta \Omega_1$. This model also describes the situation of imperfect measurements, where  variable $\Omega_1'$ is measured instead of $\Omega_1$. In these cases the peaks' width in the measured variable is broadened by an additive factor $\sigma \sin \theta$ and the peak spacing is re-scaled to $2\Omega_o\cos\theta$ (for details and a figure, see \cite{SM}). This effective width and peak-spacing can be seen as errors that do not significantly affect the code if the cumulated change in peak spacing $\Omega_o(1-\cos \theta)(1/(2\Omega_o \Delta)$ lies within the peak's half-width $\Delta/2$, or $(1-\cos \theta) \lesssim \Delta^2 \ll \Omega_o T_o/\pi = 1$. Otherwise, we can adapt the code to the modified interspacing $2\Omega_o\cos\theta$ (where $\theta$ is known), with an associated error probability ${\cal E}(\sigma \tan \theta/\Omega_o)$ in frequency and ${\cal E}(\tan \theta/(\sigma T_o))$ in time, leading to $\tan \theta \ll {\rm min} \{ \pi \sigma/(2\Omega_o),\Omega_o/(2\sigma) \}$  \cite{SM}. 

Finally, two $n$ photon TF GKP states can be entangled by applying frequency CNOT gates $\hat C_{i,j}=e^{i\hat \omega_i\otimes \hat t_j}$ \cite{PhysRevA.102.012607, PhysRevA.105.052429}, implementing $\hat C_{i,j}\ket{\omega_i}_i\ket{\omega_j}_j=\ket{\omega_i}_i\ket {\omega_i+\omega_j}_j$. We define $\hat {\cal D}_{1,2} = \otimes_{i=1}^n \hat C_{(i,1),(i,2)}$ \cite{SM}, where $(i,j)$ denotes the $i$-th spatial mode of the $j$-th  TF GKP qubit, with $j=1(2)$ for the control (target) qubit. Hence, $\hat {\cal D}_{1,2} \ket{\bar k_1}_1\ket{\bar k_2}_2=\ket{\overline k_1}_1\ket{\overline{ (k_2+k_1)~{\rm mod} ~2}}_2$, and using $\ket{\bar +}_i=\frac{1}{\sqrt{2}}\left (\ket{\overline 0}_{i}+\ket{\overline 1}_{i}\right )$, we obtain $\hat {\cal D}_{1,2} \ket{\overline +}_1\ket{\overline 0}_2=1/\sqrt{2}\left (\ket{\overline 0}_1\ket{\overline 0}_2+\ket{\overline 1}_1\ket{\overline 1}_2\right )$. Interestingly, it is also possible to implement the CNOT  gate between two TF GKP qubits by coupling only {\it one} photon from the target qubit to {\it one} photon of the control qubit: since information is encoded in the collective variables $\Omega_1$ of each qubit, for qubit states as \eqref{GKP}, for instance, we also have that $e^{i n\hat \omega_{(i,1)}\otimes \hat t_{(i,2)}}\ket{\bar k_1}_1\ket{\bar k_2}_2=\ket{\overline k_1}_1\ket{\overline{ (k_2+k_1)~{\rm mod} ~2}}_2$ (see \cite{SM} for details). In \cite{osti_1502569, Jeannic} frequency controlled two-photon gates were experimentally implemented and promising proposals exist for cavity QED platforms, {\it e.g.} \cite{Simone}. 

We can compare our results to other encodings based on GKP-like states within the continuous modes of single photons. In the TF domain, it is possible to define GKP qubits using a frequency comb spectral distribution in single photons. In this case, one photon corresponds to one qubit, and the spectral function of each photon defines a two-level-like system that exhibits local robustness against TF displacements \cite{PhysRevA.102.012607, Outro}. This is a single-mode classical-like effect independent of the number of photons involved, and if the photon's peaks are separated by $2\Omega_o/\sqrt{n}$ (as in \eqref{GKP}), the protection against displacement errors is limited to amplitudes $\delta_{\omega_j} \sim \Omega_o/(2\sqrt{n})$ {\it per} photon. Of course, it's possible to enhance protection against local (single-photon) quibit flipping and dephasing by using entangled photons. However, the so constructed codes operate similarly to discrete ones \cite{Shor,Steane,5qubits,gottesman1997stabilizer}, and correction for qubit flip and dephasing requires using entangled states of at least five photons. In contrast, the encoding proposed here demonstrates enhanced protection starting from $n=2$. 

Based on our analysis, we can reinterpret the results of \cite{PhysRevA.102.012607} as the production of a two-photon GKP state. While it remains a GKP state on a small scale, the existing techniques for producing QB GKP states involve the manipulation of Schrödinger kittens with a mean photon number of the order of one. Therefore, TF GKP states hold significant promise:  techniques to directly generate large entangled states \cite{Rempe, PhysRevLett.121.250505,doi:10.1126/science.aah4758,Yang:2021wrh,Corrielli:23,AndreaSabattoli:22, unknown, Liscidini:19} and high dimensional combs \cite{PhysRevLett.123.193603}, to manipulate photons using non-linear devices so as to implement single mode \cite{PhysRevLett.130.240801,Kurzyna:22,lipka2023ultrafast} or two-mode controlled \cite{Jeannic, osti_1502569, Simone} time-frequency operators in the universal set rapidly develop, together with high performance frequency (or mode) resolved \cite{PhysRevA.102.053707,PhysRevLett.120.030502,Luo:10} and non-destructive single-photon detectors \cite{NonDestructive, PhysRevLett.126.253603}. 

In conclusion, we have introduced and conducted an extensive study on a novel quantum optical encoding method for GKP states that can be implemented in small scale in many laboratories using current technology. Using the presently available TF GKP states, we can already envision applications in various domains, as quantum communications \cite{SaraEleni, fukui_all-optical_2021}, quantum computation \cite{KLM}, and quantum metrology \cite{Terhal}. Moreover, we can broaden the scope of potential applications by applying the proposed GKP encoding from the flourishing domain of TF based quantum photonics \cite{Lu:23} to other continuous degrees of freedom of mode entangled photons, such as their transverse position and momentum \cite{PhysRevA.95.042311, tasca_continuous_2011}, the propagation direction \cite{PhysRevX.4.031007,Raymond:23} and even to individual electrons in distinguishable modes with an underlying bosonic structure \cite{PRXQuantum.2.020314}.  An interesting perspective involves adapting recent theoretical and experimental advancements related to QB GKP states in quantum information to the modal domain \cite{calcluth2023sufficient, Calcluth2022efficientsimulation,PhysRevLett.130.090602, bourassa_blueprint_2021,PhysRevLett.112.120504, PhysRevA.102.062411,PhysRevA.99.032344, app13169462}. Finally, the tools we provide for defining quantum continuous variables using collective variables of single photons can also be extended to other error-correcting codes, such as cat codes \cite{guillaud_repetition_2019, PhysRevResearch.4.043065,PhysRevA.94.042332,PhysRevLett.119.030502,PhysRevLett.111.120501}. However, this remains a subject for future investigation.

\section*{Acknowledgements}

We acknowledge funding from the Plan France 2030 through the project ANR-22-PETQ-0006 and N. Fabre, F. Baboux and S. Ducci for fruitful discussions.

\onecolumngrid

\appendix

\section{Imperfect state preparation or measurement and the transverse spectral width}

We now discuss the role of a finite frequency width in the other collective variables $\Omega_{j\neq 1}$ and measurement or state preparation imperfections.  For this, we start with the general form
\be\label{entangled2}
\ket{\psi}=\int d\Omega_1...d\Omega_n F_k(\Omega_1)\Pi_{i> 1}^nG_i(\Omega_i)\ket{\omega_1,...,\omega_n},
\ee
For simplicity, we will consider $\Omega_\perp$ to be one of the variables $\Omega_{j>1}$, and we'll ignore all others, effectively considering a state with spectral amplitude $F_k(\Omega_1)G(\Omega_\perp)$. In this case, Eq. \eqref{entangled2} becomes
\be\label{entangled3}
\ket{\psi}=\int d\Omega_1 d\Omega_{\perp}F_k(\Omega_1)G(\Omega_{\perp})\ket{\omega_1,...,\omega_n}.
\ee
There are different ways to express both variables $\Omega_1$ and $\Omega_{\perp}$ according to the measured protocol at hand. In the present contribution, we considered that TF GKP states are perfectly prepared in variable $\Omega_1$ and function $G$ is a Gaussian spectral width, so $ F_k(\Omega_1)G(\Omega_{\perp})= (\sum_{s=-\infty}^{\infty} \delta(\Omega_1-(2s+k)\Omega_o))e^{-\frac{(\Omega_{\perp}-\omega_{\perp}^o)^2}{\sigma^2}}$. We considered for simplicity that the TF GKP states are perfect in variable $\Omega_1$, but we can express this variable using different variables as, for instance, $\Omega_1=\cos\theta \Omega_1'-\sin\theta \Omega_{\perp}'$, and $\Omega_{\perp}=\cos\theta \Omega_{\perp}'+\sin \theta \Omega_1'$. State \eqref{entangled3} can thus be re-expressed as 
\be\label{entangled4}
\ket{\psi}=\int d\Omega_1'd\Omega_{\perp}' F_k(\cos\theta \Omega_1'-\sin\theta \Omega_{\perp}')G(\cos\theta \Omega_{\perp}'+\sin \theta \Omega_1')\ket{\omega_1,...,\omega_n}.
\ee
Consider that we have wrongly prepared a TF GKP, meaning that instead of preparing state \eqref{entangled3} or, equivalently, state \eqref{entangled4}, we have prepared a TF GKP state in variable, say $\Omega_1'$, is equivalent to considering that we have correctly prepared the TF GKP but we're measuring it in a wrong basis, say, $\Omega_1'$. Using \eqref{entangled4} we'll rather consider this case. We can write 
\be\label{effective}
F_k(\cos\theta \Omega_1'-\sin\theta \Omega_{\perp}')G(\cos\theta \Omega_{\perp}'+\sin \theta \Omega_1')=\sum_{s=-\infty}^{\infty} \delta(\cos\theta \Omega_1'-\sin\theta \Omega_{\perp}'-(2s+k)\Omega_o)e^{-\frac{(\cos\theta \Omega_{\perp}'+\sin \theta \Omega_1'-\omega_{\perp}^o)^2}{\sigma^2}} 
\ee
For $\theta=-\pi/2$ (according to the chosen reference frame), we have that the measured states is a Gaussian of width $\sigma$, corresponding to the original distribution in variable $\Omega_{\perp}$. In order to have a better physical insight of the meaning of this expression, we'll consider $\omega_{\perp}^o=0$ to simplify the discussion but this constant can be inserted with no aditional difficulty. Indeed, state \eqref{effective} can be seen as a perfect TF GKP states but with a displacement $\sin\theta \Omega_{\perp}'/\cos\theta$ that is distributed according to a Gaussian of width $\sigma$. This displacement can be seen as an error, and since the variables are rotated, the widths of the error probability depend on the original variables and on the angle of rotation. Thus, the distribution of amplitudes of the displacement $\Omega_{\perp}'$ is such that it transforms the delta function into a distribution of width $\sigma \sin\theta$, as we can see by analyzing it for instance at point $\Omega_{\perp}'= -\sin \theta \Omega_1'/\cos\theta \pm \sigma/\cos\theta $. We have then, from the delta function, that $\cos\theta\Omega_1' = (2s+k)\Omega_o - \sin^2\theta\Omega_1'/\cos\theta \pm \sigma\sin\theta/\cos\theta$, or $\Omega_1'= (2s+k)\Omega_o\cos\theta \pm \sigma \sin\theta$.

\begin{figure}[h]
    \includegraphics[width=6cm]{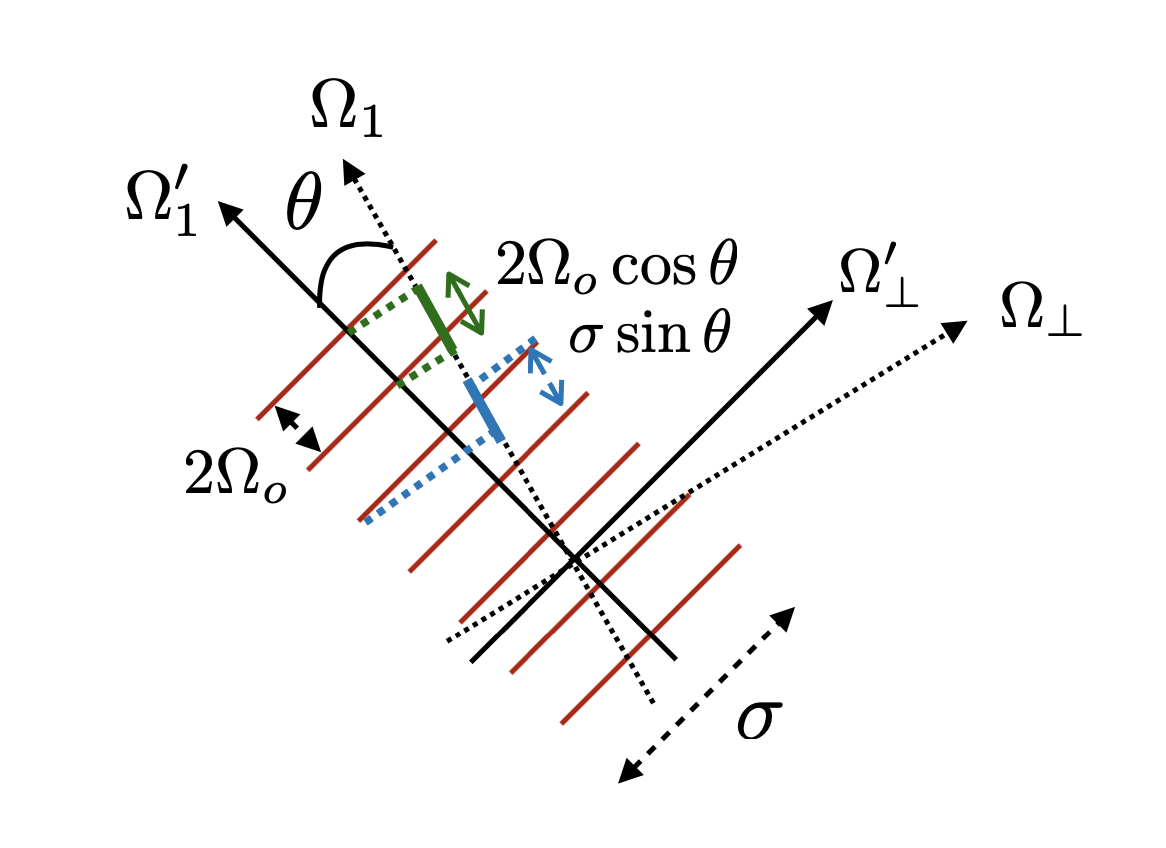}
    \caption{Effects of imperfect state preparation (or measurement) in the rotated variables $\Omega_1'$ and $\Omega_{\perp}'$. The TF GKP state in variable $\Omega_1'$ is represented by the red lines. The peak's width are broadened by a factor $\sin \theta \sigma$ and the peak interspacing becomes $2\Omega_o\cos\theta$ in $\Omega_1$.}
  \label{fig2}
\end{figure}

We have then a re-scaling of the distance between peaks and as well as the increasing of the effective peak's width (see Fig. \ref{fig2}). Since now $\sigma \sin \theta$ plays the role of an effective width for $\theta \neq n\pi$, we need to keep $\sin \theta \sigma \ll \Omega_o \cos\theta/4$,  or equivalently, $\tan \theta \ll \Omega_o/(4\sigma)$, otherwise the code will have a too large error rate. On the other hand, if we want that the change of the effective peak distance keeps the state in the code space, we need to have that $\Omega_o(1-\cos\theta) \ll \sigma \sin\theta$, that leads us to $\tan \theta \ll \sin \theta/(1-\cos\theta)$. Of course, we can add the spectral width $\Delta$ to the discussion, but this will not change the results qualitatively. 

If we now see the effect of the spectral width in the collective variable $T_1$, we will obtain a similar result: $\tilde F_k(\cos\theta T_1'-\sin\theta T_{\perp}')\tilde G(\cos\theta T_{\perp}'+\sin \theta T_1')=\sum_{s=-\infty}^{\infty} e^{-\sigma^2\frac{(T_1'+t_{\perp}^o\sin\theta-(2s+k)T_o\pi\cos\theta)^2}{\sin^2\theta}} $, where $\tilde F_k$ and $\tilde G$ are the Fourier transforms of $F_k$ and $G$ respectively, so the width of $\tilde G$ is $\sigma^{-1}$. We see that, in this case, the condition guaranteeing the low error rate of the TF GKP state is different from the one previously obtained. We have now that $\tan \theta \ll \sigma T_o\pi/4= \sigma \pi/(4\Omega_o)$. The conditions for the state to remain in the code are also the analogous to the previously obtained ones, and we obtain the same conditions on $\theta$. 

The obtained conditions on $\theta$ set the limit of the correctable errors of state preparation. In systems as SPDC, we have in general that $\sigma \ll \Omega_o$, so the restrictions set by the collective time variables are more restrictive than the ones on frequency measurements. In addition, errors due to state preparation are negligeble.

Notice that the type of imperfection discussed here is a good model for the effects of non-perfect separability between variables $\Omega_1$ and $\Omega_{\perp}$, as discussed for instance in \cite{Lundeen} in the context of quantum metrology.


\section{Errors and stabilizers}

We discuss in detail the error diagnosis and correction procedure based on the measurement of different stabilizers. We start by recalling some basic principles of the GKP states, that are stabilized by operators of the type $\hat X^{2m}$ and $\hat  Z^{2m}$ ($\hat X$ and $\hat Z$ are defined in the main text) where $m$ are integers. Starting from a perfect GKP state such that $F_k(\Omega_1)=\sum_{s=-\infty}^{\infty} \delta(\Omega_1-(2s+k)\Omega_o)$, we can understand the error detection and correction as follows: we suppose that a displacement $\hat D_{\omega_j}(\delta_{t_j})= e^{i\hat \omega_j \delta_{t_j}}$ occurs, so that the function $F_k(\Omega_1) \rightarrow \tilde F_k(\Omega_1) = \sum_{s=-\infty}^{\infty}\delta(\Omega_1-(2s+k)\Omega_o)e^{i((2s+k)\Omega_o\delta_{t_j}/\sqrt{n})}$. Applying the stabilizer $\hat X^{2m}$ consists of transforming $s \rightarrow s+m$ in the delta function. By doing so, one has that $\hat X^{2m}(\hat D_{\hat \omega_j}(\delta_{t_j})\ket{\bar k})=e^{-2im\Omega_o\delta_{t_j}/\sqrt{n}}(\hat D_{\hat \omega_j}(\delta_{t_j})\ket{\bar k})$, so $\hat D_{\hat \omega_j}(\delta_{t_j})\ket{\bar k}$ is no longer an eigenstate with $+1$ eigenvalue of the stabilizer and the error can be detected and corrected.

An interesting point here is that there are many ways to construct the $\hat X (\hat Z)$ operators and consequently, the stabilizers, and this comes of course from the errors they can correct for and the fact that states $\ket{\bar k}$ are equally robust to displacement of any photon $j$. We have thus, for instance, that $\hat X= e^{-i\Omega_o \hat T_1}$ has the same effect in $\ket{\overline k}$ as $e^{-i  \Omega_o \sqrt{n}\hat t_j}$, or $e^{-i  \sum_{j=1}^{n/2}\Omega_o 2 \hat t_j/\sqrt{n}}$ and many more, as long as one only considers the information contained in variable $\Omega_1$. Consequently, the stabilizers can also be constructed in many ways, and this will prove to be useful in correcting for photon losses as we'll see in the next section.

\section{Encoding information in variable $\Omega_1$} 

In this section, we detail some operations showing the equivalence between different operators acting on the code's subspace. 

We start by computing the effect of operators $\hat C_j^2=e^{i 2\eta T_o \hat \omega_j}$ and $\hat D_j^2=e^{2i\chi \Omega_o \hat t_j}$ to codewords when there is no photon loss. We have that 
\begin{eqnarray}\label{beg}
&&e^{i 2\eta T_o \hat \omega_j}\sum_s  \int d\Omega_1d\Omega_{\perp}\delta(\Omega_1-(2s+k)\Omega_o)G(\Omega_{\perp})\ket{\omega_1, ...,\omega_n}=\nonumber \\
&&\sum_s  \int d\Omega_1d\Omega_{\perp}e^{i \frac{2\eta \pi (2s+k)}{\sqrt{n}}} \delta(\Omega_1-(2s+k)\Omega_o)\tilde G(\Omega_{\perp})\ket{\omega_1, ...,\omega_n}
\end{eqnarray}
and 
\begin{eqnarray}\label{beg2}
&&e^{i 2\chi \Omega_o \hat t_j}\ket{\overline k}=\sum_s  \int d\Omega_1d\Omega_{\perp}\delta(\Omega_1-(2s+k)\Omega_o)G(\Omega_{\perp})\ket{\omega_1, ...\omega_j+2\chi\Omega_o,...,\omega_n}=\nonumber \\
&&\sum_s  \int d\Omega_1d\Omega_{\perp} \delta(\Omega_1-(2s+k)\Omega_o-2\chi \Omega_o)G(\Omega_{\perp}+\bar \Omega_{\perp})\ket{\omega_1, ...,\omega_n},
\end{eqnarray}

where $\tilde{G}$ and $\overline{\Omega}$ are defined in the main text. We now combine the effect of both operations into $\ket{\overline k}$, with $\eta$ arbitrary (an error) and $\chi = \sqrt{n}$:
\begin{eqnarray}\label{beg3}
&&e^{i 2 \Omega_o \sqrt{n}\hat t_j}e^{i 2\eta T_o \hat \omega_j}\ket{\overline k}= e^{i 2 \Omega_o \sqrt{n}\hat t_j}\sum_s  \int d\Omega_1d\Omega_{\perp}e^{i \frac{2\eta \pi (2s+k)}{\sqrt{n}}} \delta(\Omega_1-(2s+k)\Omega_o)\tilde G(\Omega_{\perp})\ket{\omega_1, ...,\omega_n}=\nonumber \\
&&\sum_s  \int d\Omega_1d\Omega_{\perp}e^{i \frac{2\eta \pi (2s+k)}{\sqrt{n}}} \delta(\Omega_1-(2s+k)\Omega_o)\tilde G(\Omega_{\perp})\ket{\omega_1,... \omega_j+2\Omega_o\sqrt{n},...,\omega_n}=\\
&& \sum_s  \int d\Omega_1d\Omega_{\perp}e^{i \frac{2\eta \pi (2s+k)}{\sqrt{n}}} \delta(\Omega_1-(2s+k)\Omega_o+2\Omega_o)\tilde G(\Omega_{\perp}+\overline \Omega_{\perp})\ket{\omega_1, ...,\omega_j,...,\omega_n} =\nonumber \\
&&\sum_s  \int d\Omega_1d\Omega_{\perp}e^{i \frac{2\eta \pi 2(s-1)+k)}{\sqrt{n}}} \delta(\Omega_1-(2s+k)\Omega_o)\tilde G(\Omega_{\perp}+\overline \Omega_{\perp})\ket{\omega_1, ...,\omega_j,...,\omega_n}=\nonumber \\
&&e^{-i \frac{4\eta \pi}{\sqrt{n}}}\sum_s  \int d\Omega_1d\Omega_{\perp}e^{i \frac{2\eta \pi (2s+k)}{\sqrt{n}}} \delta(\Omega_1-(2s+k)\Omega_o)\tilde G(\Omega_{\perp}+\overline \Omega_{\perp})\ket{\omega_1, ..., \omega_j,...,\omega_n}
\end{eqnarray}
that shows how the stabilizers using local operators are equivalent to the ones involving global operators $\hat \Omega_1$ and $\hat T_1$ if one is only interested in the variables $\Omega_1$ and $T_1$. It's also useful to compute 
\begin{eqnarray}\label{stab}
&&e^{i 2 T_o \sqrt{n}\hat \omega_j}e^{i 2\eta \Omega_o \hat t_j}\ket{\overline k}= e^{i 2 T_o \sqrt{n}\hat \omega_j}\sum_s  \int d\Omega_1d\Omega_{\perp} \delta(\Omega_1-(2s+k)\Omega_o) G(\Omega_{\perp})\ket{\omega_1, ...,\omega_j +2\eta\Omega_o,...,\omega_n}=\nonumber \\
&&\sum_s  \int d\Omega_1d\Omega_{\perp}e^{i 2 T_o \sqrt{n}\omega_j}\delta(\Omega_1-(2s+k)\Omega_o-2\frac{\Omega_o\eta}{\sqrt{n}})G(\Omega_{\perp}+\overline \Omega_{\perp})\ket{\omega_1, ..., \omega_j,...,\omega_n}=\\
&& \sum_s  \int d\Omega_1d\Omega_{\perp}e^{i 2 T_o \Omega_1}\delta(\Omega_1-(2s+k)\Omega_o-2\frac{\Omega_o\eta}{\sqrt{n}})\tilde G(\Omega_{\perp}+\overline \Omega_{\perp})\ket{\omega_1, ..., \omega_j,...,\omega_n}\nonumber \\
&&\sum_s  \int d\Omega_1d\Omega_{\perp}e^{i 2((2s+k)\pi+2\frac{\pi\eta}{\sqrt{n}})} \delta(\Omega_1-(2s+k)\Omega_o-2\frac{\Omega_o\eta}{\sqrt{n}})\tilde G(\Omega_{\perp}+\overline \Omega_{\perp})\ket{\omega_1, ..., \omega_j,...,\omega_n}=\nonumber \\
&&e^{i \frac{4\eta \pi}{\sqrt{n}}}\sum_s  \int d\Omega_1d\Omega_{\perp} \delta(\Omega_1-(2s+k)\Omega_o-2\frac{\Omega_o\eta}{\sqrt{n}})\tilde G(\Omega_{\perp}+\overline \Omega_{\perp})\ket{\omega_1, ..., \omega_j,...,\omega_n}.
\end{eqnarray}

These relations can be directly applied to define the stabilizers detecting the photon loss. 

\section{Entangling gates}

Entanglement between TF GKP states can be created using the tools introduced in \cite{PhysRevA.105.052429}, and more specifically, the time-frequency (TF) version of the continuous variable CNOT  gate $\hat C_{(i,1),(j,2)}\ket{\omega_{i}}_{(i,1)}\ket{\omega_{j}}_{(j,2)}=\ket{\omega_{i}}_{(i,1)}\ket {\omega_i+\omega_j}_{(j,2)}$, with $\hat C_{(i,1),(j,2)}=e^{i\hat \omega_{(i,1)}\otimes \hat t_{(j,2)}}$. We define $\hat {\cal D}_{1,2} = \otimes_{i=1}^n \hat C_{(i,1),(i,2)}$, that couples the photon in the $i$-th spatial mode of the TF GKP qubit $1$ (control qubit) to the photon in the $i$-th spatial mode of the TF GKP qubit $2$ (target qubit). Notice that modes $(i,1)$ and $(i,2)$ do not overlap. Taking two TF GKP states each characterized by a collective variable $\Omega_1$ and $\Omega_2$, the action of the gate $\hat {\cal D}_{1,2}$ on two TF GKP states with $n$ photons each is given by: 
\begin{eqnarray}\label{entangle}
&&\hat {\cal D}_{1,2}\ket{\overline k_1}_1\ket{\overline k_2}_2= \hat {\cal D}_{1,2}\left (\int d\Omega_1...d\Omega_n \sum_{s_1=-\infty}^{s_1=\infty}\delta(\Omega_1-(2s_1+k_1)\Omega_o)\Pi_{i>1}^nG_i(\Omega_i)\ket{\omega_1}_{(1,1)}...\ket{\omega_n}_{(n,1)}\right )\times \nonumber \\
&&\left (\int d\Omega'_1...d\Omega'_n \sum_{s_2=-\infty}^{s_2=\infty}\delta(\Omega'_1-(2s_2+k_2)\Omega_o)\Pi_{i>1}^nG'_i(\Omega'_i)\ket{\omega'_1}_{(1,2)}...\ket{\omega'_n}_{(n,2)}\right )=\nonumber \\
&&\left (\int d\Omega_1...d\Omega_n \sum_{s_1=-\infty}^{s_1=\infty}\delta(\Omega_1-(2s_1+k_1)\Omega_o)\Pi_{i>1}^nG_i(\Omega_i)\ket{\omega_1}_{(1,1)}...\ket{\omega_n}_{(n,1)}\right )\times \nonumber \\
&&\left (\int d\Omega'_1...d\Omega'_n \sum_{s_2=-\infty}^{s_2=\infty}\delta(\Omega'_1-\Omega_1-(2s_2+k_2)\Omega_o)\Pi_{i>1}^nG'_i(\Omega'_i-\Omega_i)\ket{\omega'_1}_{(1,2)}...\ket{\omega'_n}_{(n,2)}\right ),
\end{eqnarray}
that explicits the dependency of the logical value of the target qubit $2$ on the control's one (qubit $1$) through a controlled change of parity. 

In order to illustrate the working principles of the TF GKP CNOT gate with a simple example, we can study its application to states as the one presented in Eq. (3) of the main text. We have that 
\begin{eqnarray}\label{cnot}
&&\hat {\cal D}_{1,2} \ket{\overline k_1}_1\ket{\overline k_2}_2=  \\
&&\sum_{s_1,s=-\infty}^{\infty}\bigotimes_{m=1}^{n}\ket{\frac{2s_1+k_1}{\sqrt{n}}\Omega_o}_{(m,1)}\bigotimes_{p=1}^{n}\ket{\frac{2s+k_1+k_2}{\sqrt{n}}\Omega_o}_{(p,2)} = \ket{\overline k_1}_1\ket{\overline{ (k_2+k_1)~{\rm mod} ~2}}_2, \nonumber
\end{eqnarray}
where we sum over the dummy variable $s=s_1+s_2$. Using Eq. \eqref{cnot}, we can reproduce for the TF GKP qubits the CNOT truth table. In particular, we find that entanglement can be created by defining $\ket{\overline +}_i=\frac{1}{\sqrt{2}}\left (\ket{\overline 0}_{i}+\ket{\overline 1}_{i}\right )$, $i=1,2$, and using that $\hat {\cal D}_{1,2} \ket{\overline +}_1\ket{\overline 0}_2=1/\sqrt{2}\left (\ket{\overline 0}_1\ket{\overline 0}_2+\ket{\overline 1}_1\ket{\overline 1}_2\right )$.

An experiment implementing this type of controlled mode transformation was performed in \cite{osti_1502569} with frequency-bin encoded qubits. The same type of controlled interaction implemented in  \cite{osti_1502569},  where the displacement of the frequency of one photon is controlled by the frequency of the other would enable the implementation of the CNOT gate between TF GKP qubits. In this reference, electro-optical modulation of photons and a frequency controlled switch are combined and interference effects lead to implementation of the gate.  In \cite{Jeannic}, another technique was used to create entanglement between photons, which correspond to the TF beam-splitter like interaction $e^{i\frac{\pi}{4}(\hat \omega_1\otimes \hat t_2+\hat \omega_2\otimes \hat t_1)}$. Finally, in \cite{Simone}, the authors also propose a frequency dependent interaction in a cavity QED set-up relying on the scattering of photons by a quantum emitter. 

\begin{figure}[h]
    \includegraphics[width=6cm]{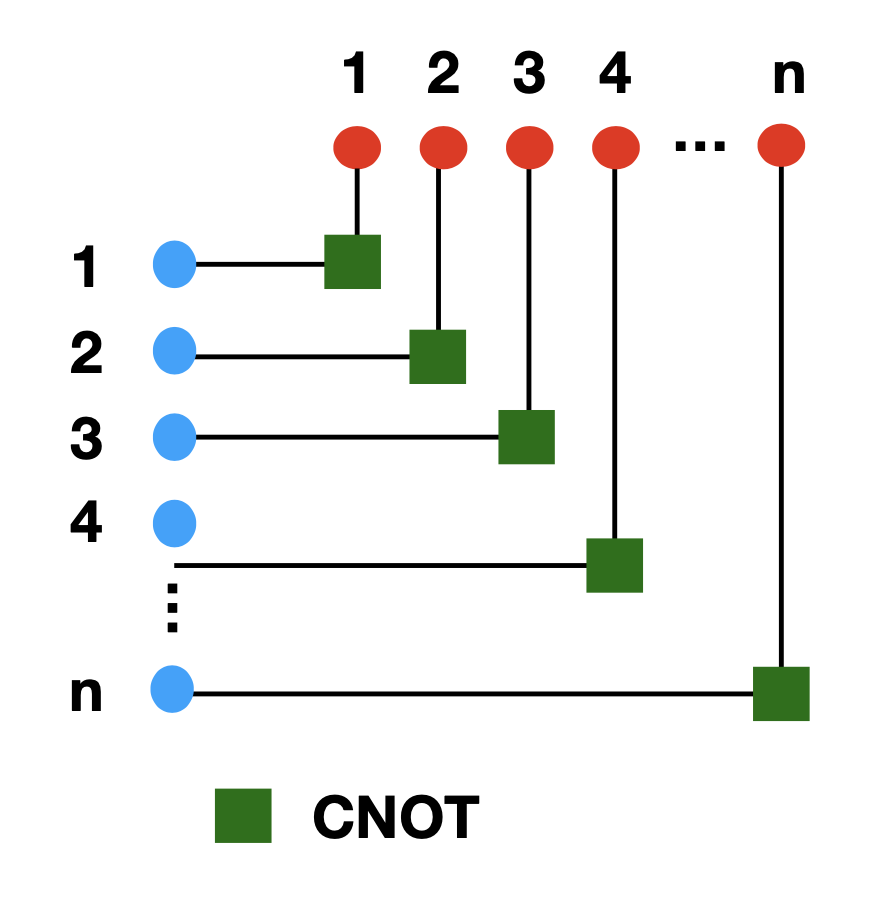}
    \caption{Representation of a CNOT gate $\hat {\cal D}_{1,2}$ between two TF GKP (in blue and in red) states comprising $n$ photons distributed in $n$ spatial modes each. Photons (represented by red and blue circles) interact two by two, and each photon pair interact only once {\it via} a TF CNOT gate $\hat C_{i,i}$ (green box). Different combinations of photons in spatial modes are possible (corresponding to different bijections $j$), the one represented in the figure is an arbitrary choice of pairing that simplifies the notation and the presentation.}
  \label{figGate}
\end{figure}

Notice that in order to entangle two TF GKP states we only require photons to interact two by two, and that we have a freedom of choice to determine which pairs interact. An example of a circuit involving two $n$ photon TF GKP states is shown in Fig. \ref{figGate}.

Finally, it would also be possible to implement a CNOT TF GKP gate using the interaction between only two photons. This is a state dependent procedure (it depends on the function $G$) but in the case of states as (3) in the main text, for instance, the conditional displacement of the TF GKP state can be implemented using only two photons. Indeed, using, for instance, $\hat {\cal C}_{1_1,1_2} = {\rm exp}(i n \hat \omega_{(1,1)}\otimes \hat t_{(1,2)} )$ (gate that couples mode $1$ of both TF GKP states), so that $\hat {\cal C}_{(1,1),(1,2)} \ket{\frac{2s_1+k_1}{\sqrt{n}}\Omega_o}_{(1,1)}\ket{\frac{2s_2+k_2}{\sqrt{n}}\Omega_o}_{(1,2)}=\ket{\frac{2s_1+k_1}{\sqrt{n}}\Omega_o}_{(1,1)}\ket{\left (\frac{2s_2+k_2}{\sqrt{n}}+(2s_1+k_1)\sqrt{n}\right )\Omega_o}_{(1,2)}$. This means that a change of variables $\Omega'_1 \rightarrow \Omega'_1 -2s_1-k_1$ can be performed, which is exactly the same change of variables that is performed when gate $\hat {\cal D}_{1,2}$ is applied to states of the type $\ket{\frac{2s_1+k_1}{\sqrt{n}}\Omega_o}_{(1,1)}\ket{\frac{2s_2+k_2}{\sqrt{n}}\Omega_o}_{(1,2)}$. Hence, the effect of both gates is the same in the global variable of the target TF GKP qubit. This geometrical effect is illustrated in Fig. 1 of the main text. 

\section{The effects of photon loss: adapting and restoring}

 There are two ways to cope with photon losses. We can adapt the code to a different effective peak interspacing, as mentioned in the main text, for perfect GKP states (and the other states involved in the computation should follow the same procedure). Alternatively, in the case of physical TF GKP states, we can consider that such effective modification interspacing does not significantly affect the code if the conditions states in the main text are observed, {\it i.e.}, if the cumulated effective peak interspacing modification $T_om/n$ in the whole state, that contains $\approx 1/(2\Delta T_o)$ peaks,  is within each peak's half-width $\Delta$. This leads to the condition $2T_o m/n\times 1/(2\Delta T_o) = m/(\Delta n) \leq \Delta$, or $n/m\geq 1/\Delta^2 \gg \pi/(\Omega_o T_o)=1$. These errors propagate with the number of gates, and we provide here a simple illustrative example of how this can be considered and corrected for. 
 
 We can for instance analyze $\hat Z\ket{\tilde 0}_{-1}$:
 \begin{eqnarray}\label{photonloss}
&&\hat Z \int \hat a_1(\omega)d\omega\ket{\tilde k}=\int d\Omega_1...d\Omega_n \tilde F_k (\Omega_1)\Pi_{i>1}^nG_i(\Omega_i)\ket{0,\omega_2-\frac{\Omega_o}{\sqrt{n}},...,\omega_n-\frac{\Omega_o}{\sqrt{n}} }= \nonumber \\
&&\int d\Omega_1...d\Omega_n \tilde F_k (\Omega_1+\frac{\Omega_o(n-1)}{n})\Pi_{i>1}^n G_i(\Omega_i')\ket{0,\omega_2,...,\omega_n}=\\
&&\int d\Omega_1...d\Omega_n \tilde F_{(k+1){\rm mod} ~2} (\Omega_1-\frac{\Omega_o}{n})\Pi_{i>1}^n G_i(\Omega_i')\ket{0,\omega_2,...,\omega_n}\nonumber.
\end{eqnarray}
We can see that using $\Omega_o'=\Omega_on/(n-1)$ in the definition of $\hat Z$ would lead to the transformation $\hat Z \ket{\tilde k}_{-1}=\ket{\tilde k}_{-1}$. However, the obtained state \eqref{photonloss} can also be seen as a displaced TF GKP state, and it is easy to verify that if $m$ photons are lost we obtain 
\be\label{m}
\hat Z \ket{\tilde k}_{-m}=\int d\Omega_1...d\Omega_n \tilde F_{(k+1){\rm mod} ~2} (\Omega_1-\frac{m\Omega_o}{n})\Pi_{i>1}^n G_i(\Omega_i')\ket{0,0,..\omega_{m+1},...,\omega_n}
\ee
and also that 
\be \label{repet}
\hat Z^m \ket{\tilde k}_{-1}=\int d\Omega_1...d\Omega_n \tilde F_{(k+1){\rm mod} ~2} (\Omega_1-\frac{m\Omega_o}{n})\Pi_{i>1}^n G_i(\Omega_i')\ket{0,\omega_2,...,\omega_n}
\ee

So, photon losses can be tolerated if the associated displacement do not significantly affect the code, and it can be corrected for by applying displacement operators after the implementation of $m$ gates if $m$ is such that the cumulated error leads to intolerable discrepancy of the code. 

If one considers more complex gates, as entangling ones, errors will also be introduced by the lost photon, but can be limited if one knows the mode from which the photon that has been lost, since we can adapt operations (as entangling ones) so as not to include this mode. The best strategy is to add a new photon to the code, which can be done using entangling gates as the ones introduced in \cite{PhysRevA.105.052429} (and implemented in \cite{Jeannic}) and temporal or frequency displacements \cite{Kurzyna:22}. We can see that 

\be \label{recover}
\ket{\tilde k}_r=\hat a_1(\omega')^{\dagger}\ket{\tilde k}_{-1}=\ket{\omega'}_1 \int d\Omega_1...d\Omega_n \tilde F_k (\Omega_1)\Pi_{i>1}^n G_i(\Omega_i')\ket{\omega_2,...,\omega_n},
\ee
and that 
\be\label{condisplace}
e^{i\hat t_1 \hat \omega_2}\ket{\tilde k}_r= \int d\Omega_1...d\Omega_n \tilde F_k (\Omega_1)\Pi_{i>1}^n G_i(\Omega_i')\ket{\omega'+\omega_2, \omega_2,...,\omega_n}.
\ee
Since we have that $\omega_1=\sum_i \alpha_{j1}\Omega_j$ and  $\omega_2=\sum_i \alpha_{j2}\Omega_j$, recovering a perfect TF GKP state of $n$ photons depend on the functions $G$ as well. For instance, when they are delta functions and we have a state as Eq. (3) of the main text, we simply have to displace state in Eq. \eqref{condisplace} of $-\omega'$ in mode $1$ to recover from the photon loss. In the general case, we can think of more involved solutions but as a general message we have that for each photon loss model corresponds a most suitable code (with corresponding functions $G$). In the present case, we considered a frequency independent loss channel, and highly frequency correlated TF GKP codes are the most easily correctable for such photon losses. If we had considered a monochromatic loss channel, this type of state would be completely destroyed by a photon loss and could not be restored. One would have then to consider a TF GKP state with large temporal correlation. So, depending on the experimental situation at hand, one can choose the most suitable way to encode information in time and frequency variables so as to enable recovering from photon losses by photon addition.

\bibliography{biblioFreqGKP}

\begin{thebibliography}{101}
\bibitem [{Not({\natexlab{b}})}]{NoteHOM}%
  \BibitemOpen
  \href@noop {} {\bibinfo {title} {Notice that this is different from the
  arbitrary choice we made previously of setting $\alpha_{i,1}=1/\sqrt{n}
  ~\forall ~ i$.}} ({\natexlab{b}})\BibitemShut {NoStop}%
\bibitem [{Com()}]{CommentDisp}%
  \BibitemOpen
  \href@noop {} {\bibinfo {title} {This can be seen by comparing the error
  probability of a single photon comb-like state with interspacing
  $2\omega_o/\sqrt{n}$}}\BibitemShut {NoStop}%
\end{thebibliography}

\end{document}